\documentclass{ws-procs9x6}
\usepackage{epsfig}

\usepackage{graphicx}
\usepackage[figuresright]{rotating}

\newcommand{\lsim}{\mathrel{\mathop{\kern 0pt \rlap
  {\raise.2ex\hbox{$<$}}}
  \lower.9ex\hbox{\kern-.190em $\sim$}}}
\newcommand{\gsim}{\mathrel{\mathop{\kern 0pt \rlap
  {\raise.2ex\hbox{$>$}}}
  \lower.9ex\hbox{\kern-.190em $\sim$}}}
\newcommand{\gagamma}{g_{a\gamma\gamma}}

\newcommand{\AmS}{{\protect\the\textfont2
  A\kern-.1667em\lower.5ex\hbox{M}\kern-.125emS}}

\textfloatsep=\floatsep           
\intextsep=\floatsep

\title{The CERN Axion Solar Telescope (CAST): status and prospects}

\author{I. G. ~Irastorza$^{\lowercase{a}}$\footnote{\lowercase{Attending
speaker, E-mail: Igor.Irastorza@cern.ch}},
S.~Andriamonje$^{\lowercase{b}}$, E.~Arik$^{\lowercase{c}}$,
D.~Autiero$^{\lowercase{a}}$, F.~Avignone$^{\lowercase{d}}$,
K.~Barth$^{\lowercase{a}}$, E.~Bingol$^{\lowercase{m}}$,
H.~Brauninger$^{\lowercase{e}}$, R.~Brodzinski$^{\lowercase{f}}$,
J.~Carmona$^{\lowercase{g}}$, E.~Chesi$^{\lowercase{a,p}}$,
S.~Cebrian$^{\lowercase{g}}$, S.~Cetin$^{\lowercase{c}}$ J.
~Collar$^{\lowercase{h}}$, R.~Creswick$^{\lowercase{d}}$,
T.~Dafni$^{\lowercase{m}}$, R.~De~Oliveira$^{\lowercase{a}}$,
S.~Dedoussis$^{\lowercase{i}}$, A.~Delbart$^{\lowercase{b}}$,
L.~Di~Lella$^{\lowercase{a}}$, C.~Eleftheriadis$^{\lowercase{i}}$,
G.~Fanourakis$^{\lowercase{j}}$, H.~Farach$^{\lowercase{d}}$,
H.~Fischer$^{\lowercase{k}}$, F.~Formenti$^{\lowercase{a}}$,
T.~Geralis$^{\lowercase{j}}$, I.~Giomataris$^{\lowercase{b}}$,
S.~Gninenko$^{\lowercase{l}}$, N.~Goloubev$^{\lowercase{l}}$,
R.~Hartmann$^{\lowercase{e}}$, M.~Hasinoff$^{\lowercase{a,p}}$,
D.~Hoffmann$^{\lowercase{m}}$, J.~Jacoby$^{\lowercase{m}}$,
D.~Kang$^{\lowercase{k}}$, K.~K\"{o}nigsmann$^{\lowercase{k}}$,
R.~Kotthaus$^{\lowercase{n}}$, M.~Krcmar$^{\lowercase{o}}$,
M.~Kuster$^{\lowercase{e}}$, B.~Lakic$^{\lowercase{o}}$,
A.~Liolios$^{\lowercase{i}}$, A.~Ljubicic$^{\lowercase{o}}$,
G.~Lutz$^{\lowercase{n}}$, G.~Luzon$^{\lowercase{g}}$,
H.~Miley$^{\lowercase{f}}$, A.~Morales$^{\lowercase{g}}$,
J.~Morales$^{\lowercase{g}}$, M.~Mutterer$^{\lowercase{m}}$,
A.~Nikolaidis$^{\lowercase{i}}$, A.~Ortiz$^{\lowercase{g}}$,
T.~Papaevangelou$^{\lowercase{i}}$, A.~Placci$^{\lowercase{a}}$,
G.~Raffelt$^{\lowercase{n}}$, H. ~Riege$^{\lowercase{a}}$,
M.~Sarsa$^{\lowercase{g}}$, I.~Savvidis$^{\lowercase{i}}$,
R.~Schopper$^{\lowercase{m}}$, I.~Semertzidis$^{\lowercase{m}}$,
C.~Spano$^{\lowercase{k}}$, J.~Villar$^{\lowercase{g}}$,
B.~Vullierme$^{\lowercase{a}}$, L.~Walckiers$^{\lowercase{a}}$,
K.~Zachariadou$^{\lowercase{j}}$,
K.~Zioutas$^{\lowercase{a,i}}$\\}

\address{\scriptsize $^{\lowercase{a}}$European Organization for Nuclear Research (CERN), Geneve, Switzerland\\
$^{\lowercase{b}}$DAPNIA, Centre d'Etudes de Saclay (CEA-Saclay),
Gif-Sur-Yvette, France\\ $^{\lowercase{c}}$Department of Physics,
Bogazici University, Istambul, Turkey\\
$^{\lowercase{d}}$Department of Physics and Astronomy, U. South
Carolina, Columbia, Sc, USA\\
$^{\lowercase{e}}$Max-Planck-Institut f\"{u}r Extraterrestrische
Physik, MPG, Garching, Germany\\ $^{\lowercase{f}}$Pacific
Northwest National Laboratory, Richland, Wa, USA\\
$^{\lowercase{g}}$Instituto de F\'{\i}sica Nuclear y Altas Energ\'{\i}as,
Universidad de Zaragoza, Zaragoza, Spain\\
$^{\lowercase{h}}$Enrico Fermi Institute, University of Chicago,
Chicago, Il, USA\\ $^{\lowercase{i}}$Aristotle University of
Thessalon\'{\i}ki, Thessaloniki, Greece\\ $^{\lowercase{j}}$National
Center for Scientific Research "Demokritos" (NRCPS), Athens,
Greece\\ $^{\lowercase{k}}$Albert-Ludwigs-Universit\"{a}t Freiburg,
Freiburg, Germany\\ $^{\lowercase{l}}$Institute for Nuclear
Research (INR), Russian Academy of Sciences, Moscow, Russia\\
$^{\lowercase{m}}$Institut f\"{u}r Kernphysik, Technische Universitat
Darmstadt, Darmstadt, Germany\\
$^{\lowercase{n}}$Max-Planck-Institut f\"{u}r Physik, Munich,
Germany\\ $^{\lowercase{o}}$Ruder Boskovic Institute, Zagreb,
Croatia\\$^{\lowercase{p}}$Department of Physics and Astronomy, U.
of British Columbia, Vancouver, Canada}

\begin{document}

\maketitle

\abstracts{The CAST experiment is being mounted at CERN. It will
make use of a decommissioned LHC test magnet to look for solar
axions through its conversion into photons inside the magnetic
field. The magnet has a field of 9 Tesla and length of 10 m and is
installed in a platform which allows to move it $\pm 8 ^\circ$
vertically and $\pm 40 ^\circ$ horizontally. According to these
numbers we expect a sensitivity in axion-photon coupling $\gagamma
\lsim 5 \times 10^{-11}$ GeV$^{-1}$ for $m_a \lsim 10^{-2}$ eV,
and with a gas filled tube $\gagamma \lsim 10^{-10}$ GeV$^{-1}$
for $m_a \lsim 2$ eV.}


\section{Introduction}

Axions are light pseudoscalar particles that arise in theories in
which the Peccei-Quinn U(1) symmetry has been introduced to solve
the strong CP problem\cite{Peccei:1977hh,raffelt_review}. They
could have been produced in early stages of the Universe being
attractive candidates to the cold Dark Matter (and in some
particular scenarios to the hot Dark Matter) responsible to the
1/3 of the ingredients of a flat universe.
Axions could also be copiously produced in the core of the stars
by means of the Primakoff conversion of the plasma photons. In
particular, a nearby and powerful source of stellar axions would
be the Sun.

The solar axion flux can be easily
estimated\cite{vanBibber:1989ge,Creswick} within the standard
solar model,
resulting in an axion flux of an average
energy of about 4 keV.
These solar axions (as well as cosmological ones) have been
searched for in the past by a number of
experiments\cite{Rosenberg:wb}. In particular, there have been two
previous experiments using magnets to search for
them\cite{Lazarus:1992ry,Inoue:2002qy}, following the original
idea of \cite{Sikivie}. The CAST
collaboration\cite{Zioutas:1998cc} is using a decommissioned LHC
test magnet which improves substantially the sensitivity to solar
axions with respect to the previous experiments. In the following
we make a short description of the experiment as well as its
status and prospects.

\section{Description and status of the experiment}

\begin{figure*}[t] \centerline{ \epsfxsize=10cm
\epsffile{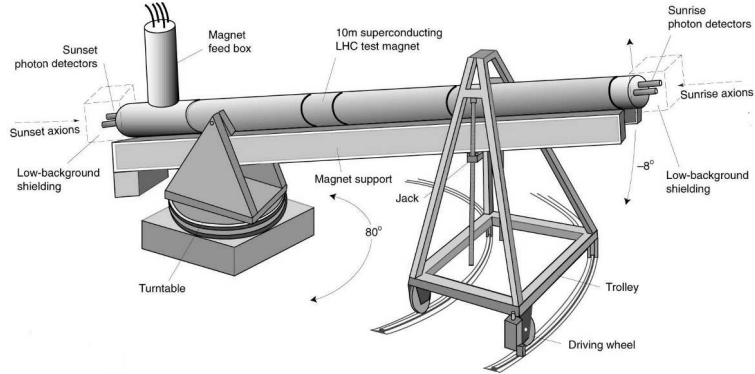} }
 \caption{Schematic view of the CAST experimental setup. The 10 m long
 LHC test magnet is mounted on a platform like the one shown in
 the drawing, allowing a movement of
$\pm 8 ^\circ$ vertically and $\pm 40 ^\circ$ horizontally. The
detectors will be located at both ends of the magnet, and will be
exposed of axion-induced X-rays during 3 hours per day in average.
}
 \label{platform}
\end{figure*}

As stated before, CAST uses a decommissioned LHC test magnet of 9
Tesla and 10 m long. The magnet is twin-aperture and has a
effective cross section of $\sim2\times14$ cm$^2$. At both ends,
several detectors will look for the X-rays originated by the
conversion of the axions inside the magnet when it is pointing to
the Sun. The magnet is mounted on a platform which allows a
movement of $\pm 8 ^\circ$ vertically and $\pm 40 ^\circ$
horizontally. With this platform it can be oriented to certain
directions and track the Sun during about three hours per day in
average (1.5 at sunrise and 1.5 at sunset). The limitation in the
movement of the magnet --and therefore in the amount of exposition
to the Sun-- is fixed by the cryogenic system that keeps it
superconducting. In Fig.~\ref{platform} an schematic view of the
experimental setup is plotted, showing the magnet and the platform
to move it. The position of the X-ray detectors at both ends of
the magnet is also marked. At the time of writing this text, the
full cryogenic installation is finished, the magnet reaching
routinely its nominal current and magnetic field. The hardware and
software of the tracking system are also fully operational, and
the orientation of the magnet has been correlated with celestial
coordinates.

The probability that an axion going through the transverse
magnetic field $B$ over a length $L$ will convert to a photon is
given by:

\begin{eqnarray}\label{conversion_prob}
  P_{a\gamma} = 2.1 \times 10^{-17} \left(\frac{B}{9 \mathrm{\ T}}\right)^2
  \left(\frac{L}{10 \mathrm{\ m}}\right)^2 \left(\gagamma \times 10^{10}
  \mathrm{\ GeV}^{-1}\right)^2\left|\mathcal{M}\right|^2
\end{eqnarray}

\noindent where the matrix element
$\left|\mathcal{M}\right|^2=2(1-\cos q L) / (qL)^2$ accounts for
the coherence of the process,
being $q$ the momentum exchange. In our case, the axion and photon
waves keep in phase ($\left|\mathcal{M}\right|^2 \simeq 1$) for
axion masses up to $\sim 10^{-2}$ eV, while for higher masses
$\left|\mathcal{M}\right|$ begins to decrease, an so does the
sensitivity of the experiment. To cope with this coherence loss, a
second phase of CAST is planned filling the beam pipe inside the
magnet with a gas to give a mass to the photons. For axion masses
that match the photon mass, the coherence is restored. By changing
the pressure of the gas inside the pipe the sensitivity of the
experiment can be extended to higher axion masses.



Three different types of detectors have been developed to be used
to detect the X-rays originated by the conversion of the axions
inside the magnet: a time projection chamber (TPC), a CCD and a
micromegas detector. The TPC will profit of the stability and
robustness of a well-known technique with position sensitivity to
single out axions coming out of the magnet bores. It is placed in
one of the ends of the magnet looking for X-rays from "sunset"
axions. In the other side of the magnet, facing "sunrise" axions,
two different detectors will be working at the same time (recall
that LHC magnets are twin aperture). One smaller gas chamber will
test the more novel micromegas principle\cite{Giomataris:1995fq}
with a potentially higher rejection factor due to the better
spatial resolution. A CCD detector or a smaller micromegas chamber
will work in conjunction with a mirror system to focus X-rays
coming out of the magnet bores. This special device is commented
later on.

The installation of the detectors in the experimental site is in
an advanced phase. Two of them (the TPC and the micromegas) are
already installed on the corresponding ends of the magnet, and
their normal operation has been checked. In fact, at the time of
writing this text, some first preliminary hours of data have been
taken with the TPC in "axion-sensitive" conditions, i.~e., with
the magnet on and tracking the Sun. The installation phase of CAST
will be fully completed when the third detector together with the
focusing mirror system will be installed in the next days.

To obtain a reasonable background level several low background
techniques are being used. The radioactivity of the components
near the detectors have been measured, in order to avoid the
presence of substantial background sources. CAST experimental site
is at surface level, so the presence of cosmic rays will be an
important component of the background. To cope with this, some
shielding is being designed. In spite of weight and space
constrains imposed by the experimental area and set-up, some
preliminary tests with the TPC in the laboratory indicate that the
background could be shielded down by one order of magnitude. Also
a good part of the background can be rejected off-line due to the
different signatures that X-rays and most of the background leave
in the detectors.
In addition,
the use of the above mentioned X-ray focusing mirror system is
meant to further improve the signal to noise ratio and so the
sensitivity of the experiment. Such a device, similar to those
used in X-ray astronomy, will be able to focus axion induced
X-rays from the magnet bore to a submillimeter spot. This will
allow us to use a very small detector (with lower expected
background) and increase the expected signal to background ratio
by about two orders of magnitude.

The background optimization being still in progress, some
preliminary test with the detectors show background levels below
10$^{-5}$ counts/keV/cm$^2$/s in the energy region of the solar
axion spectrum, which is already very close to the objective of
the proposal\cite{Zioutas:1998cc}, which foresee
a 3-$\sigma$ limit of $\gagamma \lsim 5
\times 10^{-11}$ GeV$^{-1}$ in the axion mass range where
coherence is preserved.

\begin{figure}[t] \centerline{ \epsfxsize=7.5cm
\epsffile{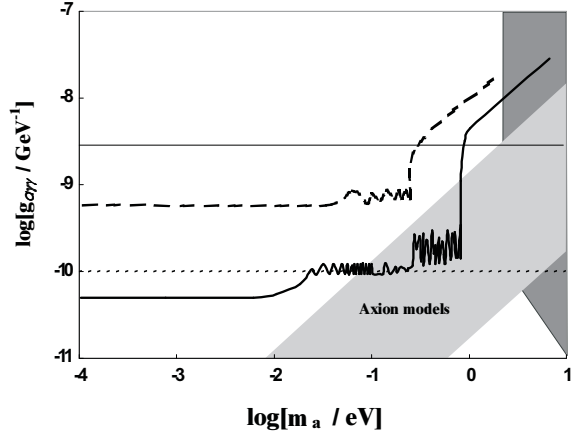} }
 \caption{Expected exclusion limit of CAST (solid line) taking
 into account both data taking phases as explained in the text.
Also shown are the limits obtained by SOLAX\protect\cite{solax}
and COSME\protect\cite{cosme} experiments using solid state
detectors (thin line), and the Tokyo
"helioscope"\protect\cite{Inoue:2002qy} (dashed line) using the
same principle as CAST. The region favoured by theoretical axion
models is also shown. Finally, the dotted line represents the
theoretical red giant bound\protect\cite{red_giant} and the grey
region on the right is excluded from the absence of an axion-decay
quasi-monochromatic photon line from galactic
clusters\protect\cite{indirect}.}
 \label{exclusion}
\end{figure}

In Fig.~\ref{exclusion} the complete expected exclusion plot is
shown, including both phases of data taking: one year in vacuum
and two more years with the tube filled with gas at varying
pressure. The first phase is responsible of the low axion mass
part of the limit ($m_a \lsim 0.02$ eV), while the second phase
will extend the limit to higher axion masses roughly to $\gagamma
\lsim 10^{-10}$ GeV$^{-1}$ for $m_a \lsim 1$ eV. From
Fig.~\ref{exclusion} one can see that CAST will be able to improve
substantially previous laboratory limits. In addition, it will go
beyond some theoretical bounds like the red giant limit. It can
also be seen that, during the second phase of data taking, CAST
will be able to enter into some theoretically favored region of
parameter space.

Finally, let's mention briefly some others goals of CAST that can
be considered as by-products of the experiment. For instance, CAST
could be able to detect the suggested monochromatic 14.4 keV
nuclear axion emission from the Sun's core\cite{Moriyama:1995bz}.
Other kinds of axion emission have been proposed in more specific
theoretical models, like Kaluza-Klein axion models, or the case
where the axion is coupled to the electron, giving rise to
axion-bremstrahlung emission. In addition, during the
non-alignment periods there will be opportunities to observe other
possible sources of axions in the sky. CAST could make some kind
of "axion astronomy" by mapping the data taken during these
periods on galactic coordinates. Examples of other possible axion
sources are the galactic center, pulsars, GRBs and supernovae.

\section{Conclusions}

The CAST experiment is now finishing its installation phase at
CERN. The 9 Tesla and 10 m long LHC test magnet is mounted in a
platform which allows to move it $\pm 8 ^\circ$ vertically and
$\pm 40 ^\circ$ horizontally. The cryogenics of the magnet and the
tracking system are ready and two gas detectors (a TPC and a
micromegas) are already installed at both ends of the magnet. Some
preliminary data have already been taken, and the definitive data
taking phase will start as soon as the third detector and the
X-ray focusing mirror system will be installed. Regarding the main
purpose of the experiment, the search for "standard" hadronic
solar axions, the preliminary tests allow us to keep the original
goal of a sensitivity to axion-photon coupling down to $\gagamma
\lsim 5 \times 10^{-11}$ GeV$^{-1}$ for $m_a \lsim 0.02$ eV. A
second phase of acquisition with a gas filled tube to provide a
mass to the photons will allow to extend a limit $\gagamma \lsim
10^{-10}$ GeV$^{-1}$ for axion masses up to $m_a \lsim 1$ eV. Some
by-products of the experiment are foreseen, like the search for
other model-specific solar axion emissions, or the exploration of
other axion sources in the sky.

\end{document}